\documentclass{PoS}

\title{A theoretical update on $t\bar{t}$ production near threshold}

\ShortTitle{$t\bar{t}$ production near threshold}

\author{\speaker{Adrian Signer}\thanks{It is a pleasure to thank
        Antonio Pineda for collaboration on the work presented here.}\\
        IPPP, Durham University (United Kingdom)\\
        E-mail: \email{adrian.signer@durham.ac.uk}}

\abstract{We present an evaluation of the $t\bar{t}$ cross section
near threshold at next-to-next-to-leading logarithmic accuracy, using
a two-step matching procedure. QED corrections are taken into account
as well and are shown to be numerically important. Finally we give an
outlook on how to further improve the theoretical predictions with
particular emphasis on how to consistently include finite width
effects.}

\FullConference{International Workshop on Top Quark Physics\\
		 January 12-15, 2006\\
		 Coimbra, Portugal}

\begin{document}

\newcommand{\MS}{$\overline{\rm MS}$}

\section{Introduction}

A future International Linear Collider (ILC) offers the opportunity to
study the top quark with unprecedented accuracy. To fully exploit the
potential of an ILC in this respect, it is essential that a dedicated
measurement of the cross section for the  production of a top antitop
quark pair close to threshold is made. Such a threshold scan allows for
an extremely precise measurement of the top-quark mass and yields
information on the top-quark width and the top-Higgs Yukawa coupling.

In order to take full advantage of future experimental data,
theoretical computations have to match the predicted experimental
precision. As is well known~\cite{FadinKhoze}, a perturbative
expansion in the strong coupling $\alpha_s$ alone is not adequate due
to the presence of terms $(\alpha_s/v)^n$ at $n$-th order in
perturbation theory, where $v\ll 1$ is the (small) velocity of the top
quarks near threshold.  These terms have to be resummed and the
theoretical computations are organized as a double expansion in the
two small parameters $\alpha_s\sim v$. Thus, a next-to-leading order
(NLO) result includes all terms that are suppressed by one power of
either $\alpha_s$ or $v$, whereas a next-to-next-to-leading order
(NNLO) result contains all terms suppressed by either $\alpha_s^2$,
$\alpha_s\, v$ or $v^2$.

Such an expansion is done most efficiently by using an effective
theory approach. From a technical point of view this amounts to using
the threshold expansion~\cite{Beneke:1997zp} to split the Feynman
diagrams into contributions due to various modes and then integrating
out unwanted modes. The modes to be considered are hard modes (with
momentum of the order of the heavy quark mass $k^\mu \sim m_t$), soft
modes (with momentum of the order of the typical momentum of the
non-relativistic top quarks $k^\mu \sim m_t\, v$), ultrasoft modes
(with momentum of the order of the typical energy of the
non-relativistic top quarks $k^\mu \sim m_t\, v^2$) and, finally,
potential modes (with $k^0 \sim m_t v^2$ and $\vec{k}\sim m_t
\vec{v}$). In a first step hard modes are integrated out, resulting in
non-relativistic QCD (NRQCD)~\cite{nrqcd}. NRQCD is matched to QCD at
a hard scale $\mu_h\sim m_t$. In a second step, soft modes and
potential gluons (and massless quarks) are integrated out. The
resulting theory is called potential NRQCD (pNRQCD)~\cite{pnrqcd} and
is matched to NRQCD at a soft scale $\mu_s\sim m_t\, v$. This theory
consists of heavy quarks with energy $E\sim m_t v^2$ and momentum
$\vec{k}\sim m_t v$, interacting through potentials (to be interpreted
as matching coefficients) and still dynamical ultrasoft gluons with
momentum $k^\mu\sim\mu_{us}\sim m_t v^2$.

This approach has been used by several groups to compute
$t\bar{t}$-pair production near threshold at NNLO (for a review
see~\cite{Hoang:2000yr}). If the cross section is expressed in terms
of a threshold mass~\cite{Beneke:1998rk,massdef} rather than the pole
mass, the position of its peak is very stable and can be predicted
with a small theoretical error. This will allow to determine the top
threshold mass and ultimately the top \MS-mass with a very small
error. The situation is much less favourable regarding the
normalization of the cross section. The corrections are huge and the
scale dependence at NNLO is larger than at NLO, indicating that this
quantity is not well under control at NNLO. 

The bad behaviour of the fixed order perturbation theory is due to the
presence of large logarithms $\log v$. In order to improve the
situation, these logarithms have to be resummed. Counting $\alpha
\log~v \sim 1$ in a NLO and NNLO calculation produces a result of
next-to-leading logarithmic (NLL) and next-next-to-leading logarithmic
(NNLL) accuracy respectively. At leading order, there are no
logarithms. Thus the leading order (LO) result coincides with the
result at leading logarithmic accuracy (LL). Within the effective
theory approach the logarithms are resummed by computing the anomalous
dimensions of the matching coefficients and solving the corresponding
renormalization group equations.

The renormalization-group improved (RGI) cross section for $t\bar{t}$
pair production close to threshold has been computed~\cite{ttrgi} a
few years ago using vNRQCD~\cite{vnrqcd}, an approach that is somewhat
different from the two step matching procedure QCD $\to$ NRQCD $\to$
pNRQCD mentioned above. In vNRQCD the relation between the soft and
ultrasoft scale is fixed from the beginning by setting $\mu_{us} =
\mu_s^2/m_t$ and there is only one step in the matching procedure, QCD
$\to$ vNRQCD. These computations showed that the $\log v$ terms are
numerically very important and improve the behaviour of the
perturbative series considerably.

So far no calculation of $t\bar{t}$ pair production at NNLL accuracy
using a pNRQCD approach has been done, even though most partial
results were available for quite a while. In Section~\ref{nnll} we
present such a computation~\cite{rgipNRQCD}. In addition we include
QED corrections which turn out to be numerically relevant. The most
important missing terms are identified and prospects for future
improvements are given in Section~\ref{beyond}. A particularly
important problem is the inclusion of effects due to the instability
of the top quarks. This will be discussed in Section~\ref{unstable}
before the conclusions are presented.

\section{Current status: $t\bar{t}$ at NNLL \label{nnll}}

The starting point for the RGI evaluation of the threshold cross
section is the fixed-order calculation presented
in~\cite{Beneke:1999qg}. The cross section obtains contributions from
$\gamma$ and $Z$ exchange. In order to simplify the discussion we
ignore the $Z$ exchange in what follows. The cross section at center
of mass energy $\sqrt{s}$ can then be written as
\begin{equation}
\sigma^\gamma (s) = \frac{4 \pi \alpha}{3\, s}\, R(s)
\label{sigma}
\end{equation}
where $\alpha$ is the electromagnetic coupling and the $R$ ratio is
expressed as a correlator of two heavy-quark vector currents $j^\mu(x)
\equiv \bar{Q} \gamma^\mu Q(x)$
\begin{equation}
R(s) = \frac{4\pi\, e_t^2}{s}\,  {\rm Im} \left[-i \int d^4 x\,  e^{i q x} 
\langle 0 | T\{j^\mu(x) j_\mu(0)\} | 0\rangle \right]
\label{Rcorr}
\end{equation}
with $q^2=s=(2m_t+E)^2$ and $e_t$ the electric charge of the top.  The
current is expressed in terms of two-component spinor fields
$\psi^\dagger$ and $\chi$ resulting in $j\sim
c_1\,\chi^\dagger\sigma^i\psi - c_2/(6m_t^2)\,
\chi^{\dagger}\sigma^i\left(i {\bf D} \right)^2 \psi + \ldots\ $.  The
contributions of the hard modes, which are to be integrated out first,
go into the coefficient functions $d_i(\mu_h)$ of the NRQCD Lagrangian
and into the matching coefficients of the current $c_i(\mu_h)$. For
most coefficients a leading-order result is sufficient, however, $c_1$
has to be computed to NNLO~\cite{c1nnlo}.  Counting $\alpha\sim
\alpha_s^2$, the one-loop exchange of a hard photon contributes at
NNLO and is taken into account by
\begin{equation}
\label{c1qed}
c_1(\mu_h) \to c_1(\mu_h) - \frac{2\, e_t^2 \alpha(\mu_h)}{\pi}
\end{equation}
Once the hard modes are integrated out, we are left with soft,
potential and ultrasoft modes. However, near threshold the only
allowed external states are potential heavy quarks and ultrasoft
gluons. Integrating out the remaining modes results in the pNRCQD
Lagrangian, which has the structure
\begin{eqnarray}
{\cal L}_{\rm pNRQCD} &=& {\cal L}^{(0)}_{\rm pNRQCD} +
\int d^3r\, \left[\psi^\dagger T^a \psi\right](\vec{x}+\vec{r})
\, \delta V\, 
\left[\chi^\dagger T^a \chi\right](\vec{x})
\label{Lpnrqcd}
\\
&+& \psi^\dagger \left(\frac{\partial^4}{8m_t^3} - 
              g_s\, \vec{x}\cdot\vec{E}\right)\psi + 
\chi^\dagger \left(-\frac{\partial^4}{8m_t^3} - 
              g_s\, \vec{x}\cdot \vec{E} \right)\chi
\nonumber
\end{eqnarray}
where the leading order Lagrangian
\begin{eqnarray}
{\cal L}^{(0)}_{\rm pNRQCD} &=& 
\psi^\dagger \left(i\partial^0 + \frac{\partial^2}{2m_t}
                   + i \Gamma_t \right)\psi + 
\chi^\dagger \left(i\partial^0 - \frac{\partial^2}{2m_t}
                   + i \Gamma_t  \right)\chi
\label{L0pnrqcd}
\\
&+& \int d^3r\, \left[\psi^\dagger T^a \psi\right](\vec{x}+\vec{r})
\left(-\frac{\alpha_s}{r}\right) 
\left[\chi^\dagger T^a \chi\right](\vec{x})
\nonumber
\end{eqnarray}
includes the leading Coulomb interaction and takes into account the
effects due to the finite width of the top quark by the shift
$E\to E+i\Gamma_t$. It is known that this does correctly take into
account the width effects at leading order, but not at NNLO, an issue
we will come back to in Section~\ref{unstable}.

The starting point is the free Green function of the Lagrangian ${\cal
L}^{(0)}_{\rm pNRQCD}$ which, from a diagramatical point of view,
includes all ladder diagrams with the exchange of an arbitrary number
of potential gluons.  The imaginary part of the Green function at the
origin is related to the $R$ ratio by
\begin{equation}
R(E) = \frac{24\pi e_t^2 N_c}{s}\, 
\left(c_1^2-c_1 c_2 \frac{E}{3\, m_t} \right) \, 
{\rm Im}\, G(0,0,E)
\label{RvsImG}
\end{equation}
where $N_c=3$ is the color factor.  In order to regularize the
ultraviolet singularity present in the Green function, we perform the
calculation using dimensional regularization in $D=d+1=4-2\epsilon$
dimensions~\cite{Beneke:1999qg}. After minimal subtraction we
obtain~\cite{Beneke:1999zr}
\begin{eqnarray}
\label{Greenorig}
G_c(0,0,E) &\equiv& \int \frac{d^d p_1}{(2\pi)^d}
\frac{d^d p_2}{(2\pi)^d}\, \tilde{G}_c(\vec{p}_1,\vec{p}_2,E)
\nonumber \\
&=& - \frac{\alpha_s\, C_F\, m_t^2}{4\pi} 
\left( \frac{1}{2\lambda} + \frac{1}{2} \ln \frac{-4 m_t E}{\mu^2}
       - \frac{1}{2} + \gamma_E + \psi(1-\lambda) \right)
\end{eqnarray}
where $\lambda\equiv C_F\, \alpha_s/(2 \sqrt{-E/m_t})$ with
$C_F=4/3$. Once the potential, $\delta V$, is known to the desired
accuracy, the NNLO corrections are calculated as single or double
insertions
\begin{equation}
\int \frac{d^d p_1}{(2\pi)^d}\frac{d^d q_1}{(2\pi)^d}
\frac{d^d q_2}{(2\pi)^d}\frac{d^d p_2}{(2\pi)^d}\,
\tilde{G}_c(\vec{p}_1,\vec{q}_1,E)\, \delta V(\vec{q}_1-\vec{q}_2)\,
\tilde{G}_c(\vec{q}_1,\vec{p}_2,E)
\end{equation}
Such insertions can result in divergences. Since the whole calculation
is performed using dimensional regularization, it is important that
the potentials, to be interpreted as matching coefficients, are
computed consistently in $D$ dimensions. Using the result for the
two-loop static potential~\cite{static}, the potential has been
computed in Ref.~\cite{Beneke:1999qg}. Here we include an additional
contribution in $\delta V$, the electromagnetic Coulomb term
\begin{equation}
\delta V \to \delta V - \frac{4\pi\, e_t^2\, \alpha}{\vec{q}^{\ 2}}
\label{EMpot}
\end{equation}
This will give rise to a NLO term $\sim \alpha/v$ from single
potential photon exchange and a NNLO term from double potential photon
exchange. We also remark that in Eq.~(\ref{Lpnrqcd}) the leading
ultrasoft interactions are included~\cite{Beneke:1999zr} even though
they only contribute beyond NNLO. They are given by the chromoelectric
dipole operator $\vec{x}\cdot \vec{E}(t,0)$ where the electric field
is understood to be multipole expanded. An additional term containing
$A^0(t,0)$ is not displayed since it can be gauged away.

We now turn to the discussion on how to resum the logarithms in this
approach. After matching NRQCD to QCD at the hard scale $\mu_h$, the
renormalization group equations are used to evolve the matching
coefficients $d_i$ from $d_i(\mu_h)$ to the soft matching scale
$d_i(\mu_s)$, thereby resumming logarithms of the form $\log
\mu_h/\mu_s$. The matching coefficients of the single heavy quark
sector can be taken from heavy quark effective
theory~\cite{shqmc}. The RGI coefficients of the four-heavy-quark
operators have been computed in Ref.~\cite{Pineda:2001ra}.

Matching pNRQCD to NRCQD at the soft scale $\mu_s$ then results in
potentials that depend on $\alpha_s(\mu_s)$ and the matching
coefficients $d_i(\mu_s)$. Again, renormalization group equations are
used to evolve the potentials down to ultrasoft scales $\mu_{us}$,
resumming $\log \mu_s/\mu_{us}$~\cite{Pineda:2001ra}. This procedure
is complicated by the fact that higher dimension operators of the
single heavy quark sector can mix through potential loops into lower
dimension operators of the heavy quark-antiquark sector. Thus, to
obtain the NLL (NNLL) running of $c_1$ the NLL (NNLL) running of (some
of the) single heavy quark operators has to be taken into account.
While all other coefficients are known to an accuracy sufficient for a
NNLL calculation, $c_1$ is only known to NLL~\cite{Pineda:2001et}.
Some NNLL terms of $c_1$ have been computed~\cite{c1nnll} but we
stress that strictly speaking the term NNLL that is commonly used to
describe the accuracy of the result is not valid.

From the discussion in the introduction it is clear that the hard,
soft and ultrasoft scales are correlated $\mu_{us}\sim \mu_s^2/\mu_h$.
As mentioned above, in the vNRQCD approach the scales are fixed by
$\mu_{us}=\mu_s^2/m_t$.  While the two-step matching procedure in
principle does allow to relax this condition, in the current approach,
the relation $\mu_{us} = \mu_s^2/\mu_h$ has been used in solving the
renormalization group equations. Thus, it is not possible to vary
$\mu_{us}$, $\mu_h$ and $\mu_{s}$ independently around their natural
values. This is somewhat unfortunate, as it would be preferable to have
a formulation where the correlation of the scales is not rigid and
which would allow to treat and vary the various matching scales
independently. This would allow to obtain a more realistic indication
of the theoretical error, in particular since some ultrasoft
logarithms at NNLL are missing in the present results. For the moment
we just have to keep this in mind when trying to assign a theoretical
error.

\begin{figure}[h]
   \medskip
   \includegraphics[width=.5\textwidth]{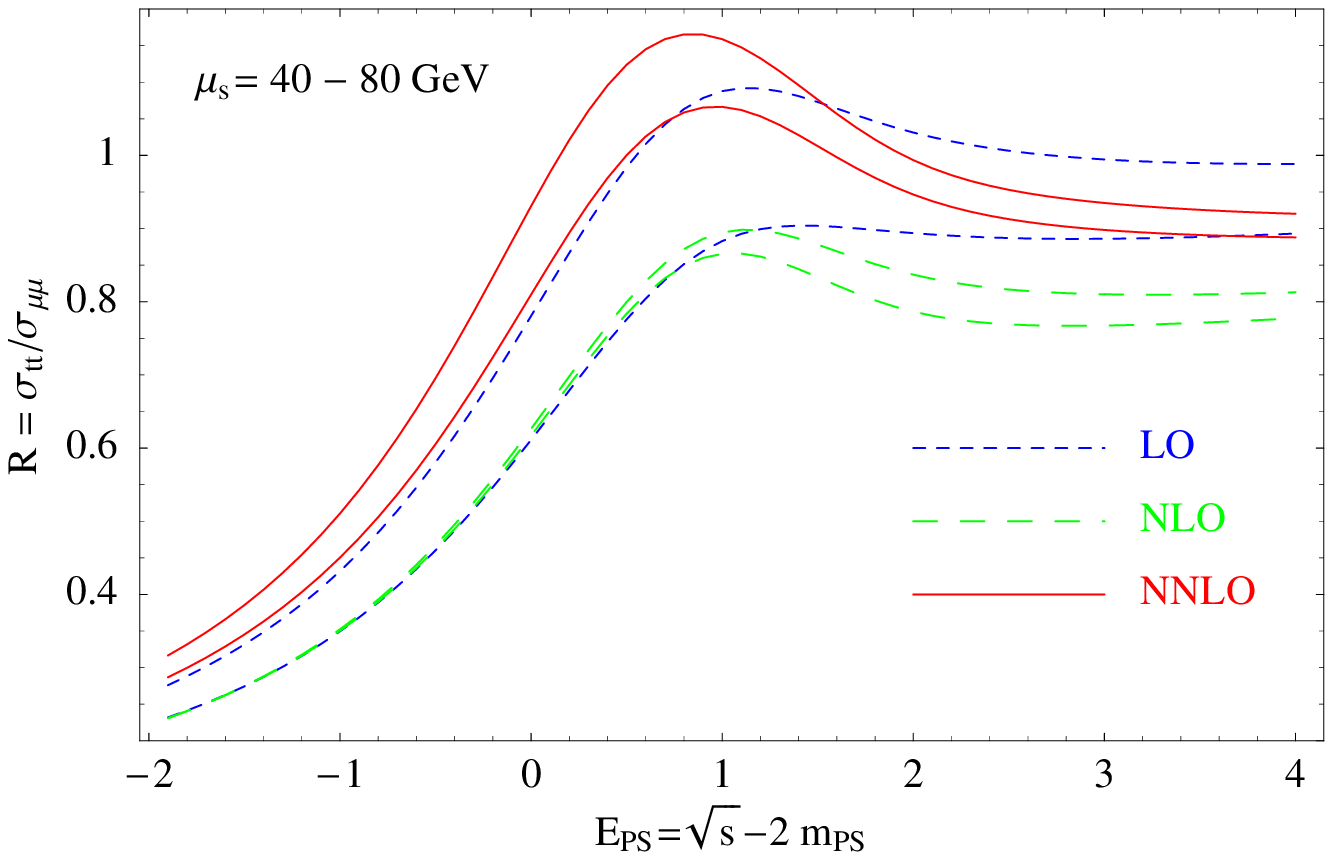}
   \includegraphics[width=.5\textwidth]{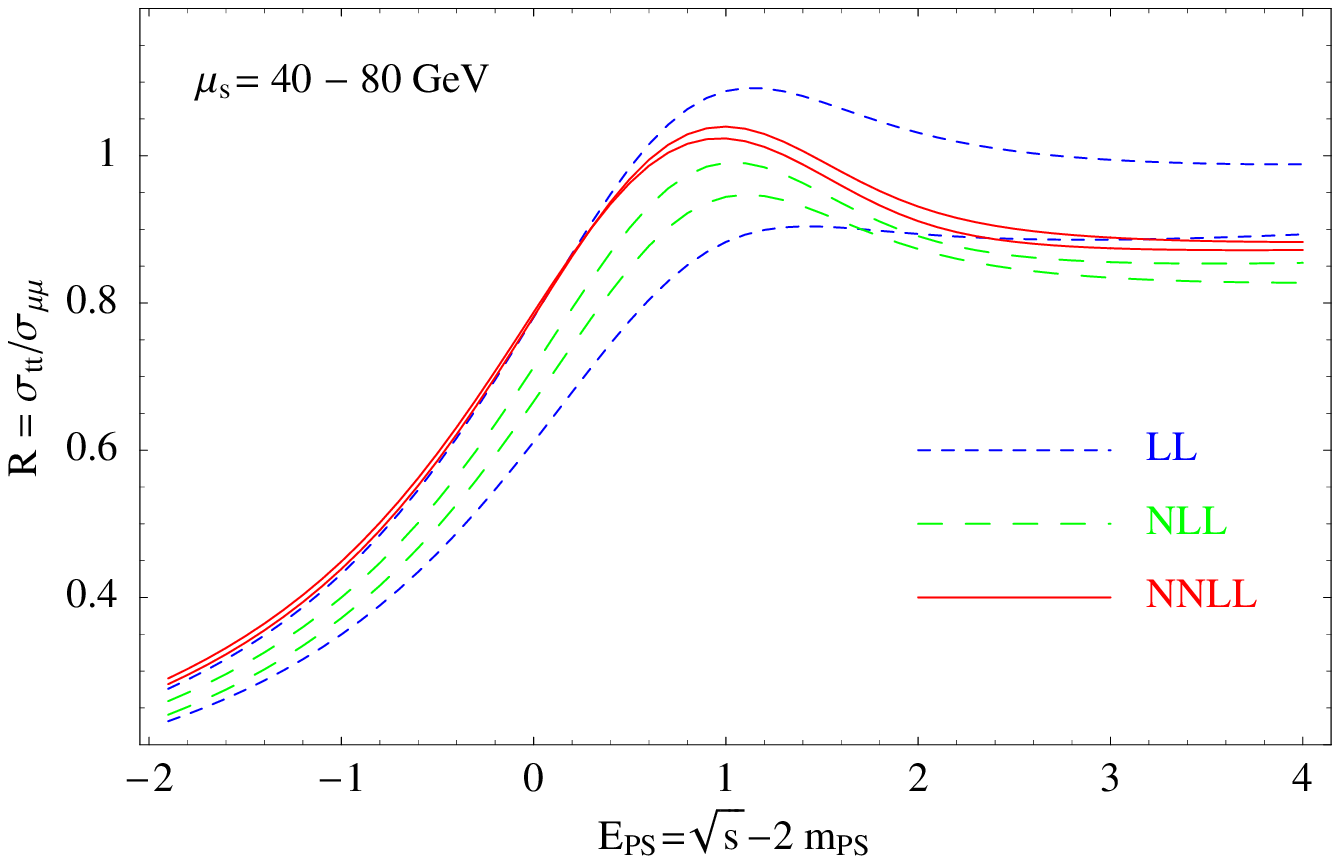}
   \caption{Threshold scan for $t\bar{t}$ using $m_{t,
   PS}=175$~GeV. The left figure shows the fixed order results, LO,
   NLO and NNLO, whereas the in right figure the RGI results LL, NLL
   and NNLL are displayed. The soft scale is varied from
   $\mu_s$=40~GeV to $\mu_s$=80~GeV }
   \label{fig:scan1}
\end{figure}

In Figure~\ref{fig:scan1} the fixed-order results are compared to the
RGI results~\cite{rgipNRQCD} obtained using the procedure described
above and using the PS mass~\cite{Beneke:1998rk}. In order to obtain a
first rough estimate of the theoretical uncertainty we show the cross
sections as bands obtained by variation of the soft scale in the
region $40~{\rm GeV} \le \mu_s \le 80~{\rm GeV}$ and setting
$\mu_h=m_t$. As was to be expected from previous results~\cite{ttrgi},
the scale dependence is much reduced once the logarithms are taken
into account. We also note that the size of the corrections decreases
for the RGI results, in particular the NNLL band is much closer to the
NNL band, and the scale dependence reduces from LL to NLL to
NNLL. However, the NNLL band does not overlap with the NLL band,
indicating that a theoretical error estimate relying on the scale
variation in this plot alone is too optimistic.

\begin{figure}[h]
   \medskip
   \includegraphics[width=.5\textwidth]{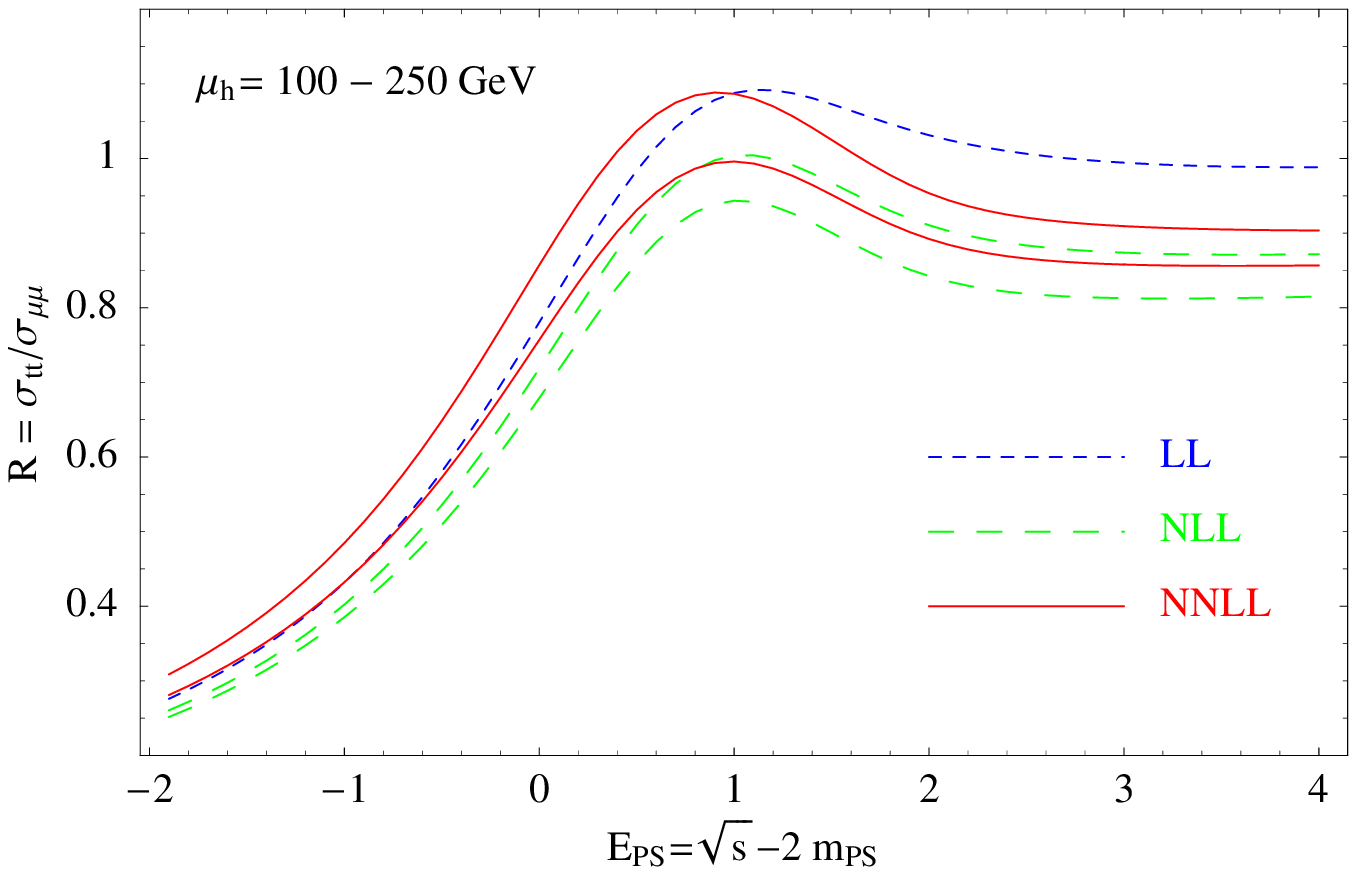}
   \includegraphics[width=.5\textwidth]{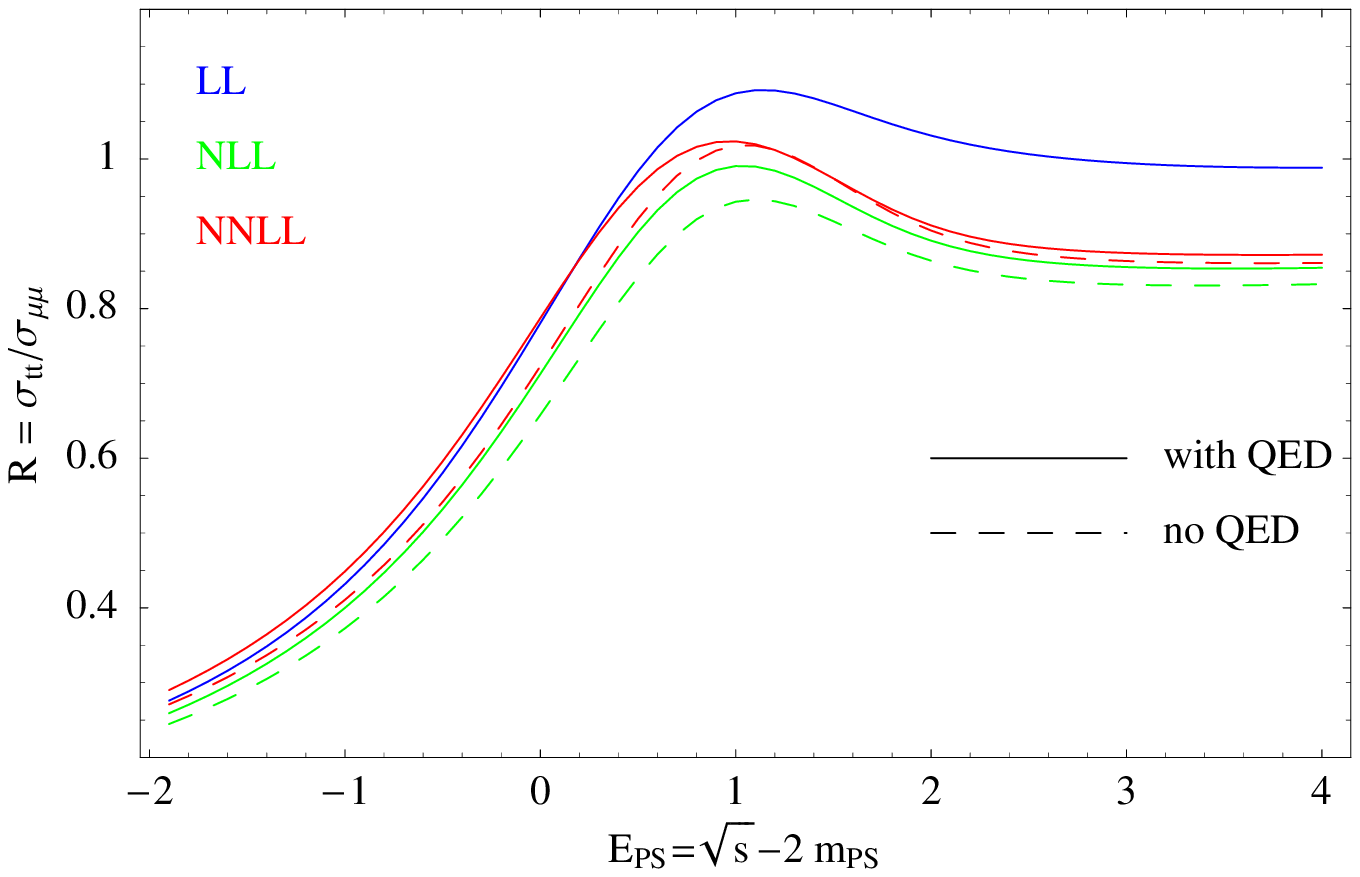}
   \caption{Left panel: Variation of the hard scale $\mu_h$ in the
   threshold scan for $m_{t, PS}=175$~GeV. The hard scale is varied
   from $100~{\rm GeV} \le \mu_h \le 250~{\rm GeV}$.  Right panel: the
   effect of including QED corrections at NLL and NNLL. There are no
   QED corrections at LL.}
   \label{fig:scan2}
\end{figure}

In the fixed-order calculations the by far dominant uncertainty came
from the variation of the soft scale. Since this problem is much less
severe after the resummation of the logarithms we have to be more
careful with other sources of uncertainties. In particular, the
missing ultrasoft contributions make it important to consider the
dependence on the other scales as well. Since $\mu_{us}$ is related to
$\mu_h$ by $\mu_{us}=\mu_s^2/\mu_h$, we consider in the left panel of
Figure~\ref{fig:scan2} the dependence on $\mu_h$ for
$\mu_s=40$~GeV. Variation of the hard scale around its natural value
$\mu_h=m_t$ by choosing $100~{\rm GeV} \le \mu_h \le 250~{\rm GeV}$
results in a scale dependence that is considerably larger than the
soft scale dependence. It is to be expected that the situation
improves once all ultrasoft logarithms at NNLL are taken into account,
but at this stage the rather large dependence of the cross section on
$\mu_h$ has to be taken into account if a theoretical error is
assigned. We also note that the dependence on $\mu_h$ only enters at
NLL, thus the LL ``band'' in the left panel of Figure~\ref{fig:scan2}
is simply a line.

Finally we turn to the right panel of Figure~\ref{fig:scan2}, where we
display the effects due to the QED corrections. As mentioned, they
enter at NLL (thus there is no effect on the LL curve) and they are
large. They change the normalization by up to 10\% and result in a
shift in the extracted $\overline{\rm MS}$ mass of up to 100~MeV,
making their inclusion imperative.

\section{Beyond NNLL \label{beyond}}

The theoretical error on the peak of the cross section and, with the
current RGI results, also its normalization is in rather good shape.
Depending on the details of the error analysis, the theoretical error
in the determination of the top mass is in the range of 50--150~MeV.
The uncertainty in the normalization of the cross section has been
quoted as 6\% in Ref~\cite{ttrgi} whereas Figure~\ref{fig:scan2} seems
to indicate a somewhat larger value. In any case it is desirable to
further reduce the theoretical error, in particular in the
determination of the normalization of the cross section, in order to
obtain the best possible measurements of the top width and maybe even
the top Yukawa coupling. 

An obvious first step is to complete the NNLL determination of the
current matching coefficient $c_1$. Once this is completed, there are
two main directions for further improvements. The first concerns the
correct implementation of finite width effects and will be discussed
in Section~\ref{unstable}. The other is the inclusion of corrections
of even higher order.

A full NNNLO calculation requires the evaluation of the matching
coefficients to the desired order, as well as the computation of the
insertions of the potentials.  Several partial results have already
been obtained. For a start, the logarithmically enhanced NNNLO terms
have been obtained a while ago~\cite{lognnnlo}. These terms are
included in the NNLL result. Furthermore, the matching coefficient for
the non-analytic $1/r^2$ potential has been computed to the required
accuracy~\cite{Kniehl:2001ju} and the NNNLO contribution due to
multiple insertions of Coulomb potentials has been
computed~\cite{Beneke:2005hg}.  However, the calculation of the static
potential and the matching coefficient $c_1$ of the current are major
obstacles on the way to a full NNNLO evaluation.

Of particular importance are the ultrasoft effects. As mentioned
before, they enter only at NNNLO and arise due to the emission and
absorption of an ultrasoft gluon through the chromoelectric dipole
operator. Thus one of the couplings is $\alpha(\mu_{us})$. Since the
ultrasoft scale is rather small, it can be argued that these
contributions are potentially particularly important NNNLO terms and
their evaluation would be an important partial result. Note that this
contribution has an ultraviolet singularity, resulting in
logarithmically enhanced NNLL terms~\cite{lognnnlo}.

The importance of NNNLO contributions can also be seen from the fact
that multiple insertions of the Coulomb potential are numerically
significant. In Figure~\ref{fig:scan1} the soft scale has not been
allowed to run below 40~GeV. The reason is that for $\mu_s \le 30$~GeV
the scale dependence get much worse. This bad behaviour is due to the
Coulomb potential and it has been shown that including multiple
insertions of the Coulomb potential the situation can be
rectified~\cite{Beneke:2005hg}.

\section{Finite width effects \label{unstable}}

The width of the top quark, $\Gamma_t$ is a leading-order effect that
has to be taken into account in the propagator of the non-relativistic
heavy quarks. This can be seen by noting that the non-relativistic
propagator $[E-\vec{p}^{\, 2}/(2 m_t)]^{-1} \sim [m_t v^2]^{-1}$
scales in the same way as the width $\Gamma_t \sim m_t \alpha_{ew}
\sim m_t v^2$. Thus the suppression due to the insertion of the self
energy is compensated by the enhancement of an additional
non-relativistic propagator and the corresponding resummation yields
\begin{equation}
\label{seresum}
\frac{1}{E-\vec{p}^{\ 2}/(2 m_t)} 
\sum_n \left(\frac{- i\, \Gamma_t}{E-\vec{p}^{\, 2}/(2 m_t)}\right)^n = 
\frac{1}{E + i\, \Gamma_t-\vec{p}^{\, 2}/(2 m_t)} 
\end{equation}
This is the justification at leading order of the replacement $E\to
E+i\, \Gamma_t$. The question is how to go about to include the finite
width effects beyond leading order. This question is actually relevant
for a large number of processes involving unstable particles, but it
is particularly pressing here, given the unprecedented precision
required in the theoretical predictions. We recall that the error of
the top quark mass $\delta m_t \sim 100$~MeV, thus  $\delta m_t \ll
\Gamma_t$. Therefore, a systematic approach on how to include finite
width effects is mandatory.

To start with we have to emphasise that the very notion of a
$t\bar{t}$ cross section breaks down. Taking into account the width
effects means we have to take into account the electroweak decay of
the top quark. Thus we are dealing with the process $\gamma^*\to
W^+\bar{b}\ W^- b$. Even if we neglect the decay of the $W$ bosons,
this opens up the possibility of many additional radiative correction
processes. There are gluons connecting the intermediate top quarks
with the decay products (non-factorizable corrections), there are
genuine electroweak corrections and, on top of this all, there are
radiative corrections due to photon exchange from any of the decay
products with any other charged particle, even the incoming
electrons. Some of the electroweak corrections have been taken into
account through absorptive parts in the NRQCD matching
conditions~\cite{Hoang:2004tg} and have been shown to be numerically
relevant. However this does not take all effects into account. In
particular QED radiative corrections which are an order $\alpha$
effect and thus contribute at NNLO have to be considered as well in a
complete solution to the problem.

The key to the solution is the observation that the situation
regarding the finite width corrections is actually very similar to
NRQCD. In QCD we have at each order $n$ of perturbation theory terms
of the form $(\alpha_s/v)^n$. Thus the suppression by the coupling is
compensated by a kinematic enhancement. This is exactly the same as in
Eq.~(\ref{seresum}). The Coulomb singularity $v=0$ has its analogy in
the propagator pole of an on-shell $E=\vec{p}^{\,2}/(2m_t)$ stable
$\Gamma_t=0$ top quark. Resummation of the $(\alpha_s/v)^n$ terms
results in a function that is well defined at $v=0$ in the same way as
the right hand side of Eq.~(\ref{seresum}) is well defined for
$E=\vec{p}^{\, 2}/(2m_t)$. In NRQCD, the problem of the Coulomb
singularity is solved by systematically expanding in both small
parameters $\alpha_s$ and $v$. Thus the basic idea is to do the same
here and use effective theory methods~\cite{Chapovsky:2001zt} to
systematically expand in both small parameters, the coupling and
$(p^2-m_t^2)/m_t^2$. 

The method has been worked out in detail and its viability has been
shown in the case of a toy model with a single resonant heavy scalar
particle~\cite{upet}. A first step towards considering pair production
near threshold, albeit for the case of $W$ bosons, has been described
in Ref.~\cite{Beneke:2004xd}. However, the main features are the same
as for top quark pair production. A consistent description requires
the introduction of additional operators and modes. The incoming
electrons and the decay products of the top quarks are to be described
by collinear modes familiar from soft-collinear effective
theory~\cite{scet}. The production and decay of the top quarks is
described by new operators $\left(\bar{e} e\right) \bar(\chi^\dagger
\psi)$, where $\bar{e}$ and $e$ are collinear electron and positron
fields respectively, and $\psi^\dagger W^+ b$ and $\chi W^- \bar{b}$,
where $W$ and $b$ describe the decay products of the top quark.
Electroweak corrections are split into various contributions. First
there are the hard corrections to be taken into account by computing
the matching coefficients of these operators to the desired
accuracy. This corresponds to the computation of the matching
coefficients in NRQCD and reproduces the so-called factorizable
corrections. We are then left with matrix element corrections due to
the still dynamical soft, potential and collinear modes which can be
associated with the non-factorizable corrections.

While there is still a long way to go to actually perform an explicit
calculation, we are confident that this will be the method of choice
to include the finite width effects in a systematic way.

\section{Conclusions}

There has been a lot of progress in the theoretical evaluation of the
$t\bar{t}$ threshold cross section and the situation concerning the
position of the peak of the cross section, relevant for the mass
measurement is very well under control. A mass measurement with an
error of 100--150~MeV is certainly achievable, as long as a threshold
mass definition is used.  Regarding the normalization of the cross
section, relevant for a measurement of the width of the top quark and
eventually its Yukawa coupling, the resummation of the logarithms has
also substantially improved the situation. However, this will need
further improvements to fully exploit the potentially very precise
experimental data. Apart from the completion of the NNLL evaluation,
the effects due to the finite width will have to be fully
understood. Obviously, any progress towards a NNNLO evaluation is most
welcome. Finally, let me conclude with an obvious but often forgotten
remark: exploiting this unique opportunity to measure some Standard
Model parameters to an unprecedented accuracy actually does require a
dedicated experimental effort in that a future International Linear
Collider will have to be run at the top threshold for a reasonable
amount of time.


\begin{thebibliography}{99}

\bibitem{FadinKhoze}
V.~S.~Fadin and V.~A.~Khoze,
JETP Lett.\  {\bf 46}, 525 (1987)
[Pisma Zh.\ Eksp.\ Teor.\ Fiz.\  {\bf 46}, 417 (1987)]; \\
V.~S.~Fadin and V.~A.~Khoze,
Sov.\ J.\ Nucl.\ Phys.\  {\bf 48}, 309 (1988)
[Yad.\ Fiz.\  {\bf 48}, 487 (1988)].

\bibitem{Beneke:1997zp}
M.~Beneke and V.~A.~Smirnov,
Nucl.\ Phys.\ B {\bf 522}, 321 (1998)
[arXiv:hep-ph/9711391].

\bibitem{nrqcd}
W.~E.~Caswell and G.~P.~Lepage,
Phys.\ Lett.\ B {\bf 167}, 437 (1986); \\
G.~T.~Bodwin, E.~Braaten and G.~P.~Lepage,
Phys.\ Rev.\ D {\bf 51}, 1125 (1995)
[Erratum-ibid.\ D {\bf 55}, 5853 (1997)]
[arXiv:hep-ph/9407339].

\bibitem{pnrqcd}
A.~Pineda and J.~Soto,
Nucl.\ Phys.\ Proc.\ Suppl.\  {\bf 64}, 428 (1998)
[arXiv:hep-ph/9707481]; \\
A.~Pineda and J.~Soto,
Phys.\ Rev.\ D {\bf 59}, 016005 (1999)
[arXiv:hep-ph/9805424].

\bibitem{Hoang:2000yr}
A.~H.~Hoang {\it et al.},
Eur.\ Phys.\ J.\ directC {\bf 2}, 1 (2000)
[arXiv:hep-ph/0001286].

\bibitem{Beneke:1998rk}
M.~Beneke,
Phys.\ Lett.\ B {\bf 434}, 115 (1998)
[arXiv:hep-ph/9804241].

\bibitem{massdef}
I.~I.~Y.~Bigi, M.~A.~Shifman and N.~Uraltsev,
Ann.\ Rev.\ Nucl.\ Part.\ Sci.\  {\bf 47}, 591 (1997)
[arXiv:hep-ph/9703290]; \\
A.~H.~Hoang, Z.~Ligeti and A.~V.~Manohar,
Phys.\ Rev.\ Lett.\  {\bf 82}, 277 (1999)
[arXiv:hep-ph/9809423]; \\
A.~Pineda,
JHEP {\bf 0106} (2001) 022
[arXiv:hep-ph/0105008].

\bibitem{ttrgi}
A.~H.~Hoang, A.~V.~Manohar, I.~W.~Stewart and T.~Teubner,
Phys.\ Rev.\ Lett.\  {\bf 86}, 1951 (2001)
[arXiv:hep-ph/0011254]; \\
A.~H.~Hoang, A.~V.~Manohar, I.~W.~Stewart and T.~Teubner,
Phys.\ Rev.\ D {\bf 65}, 014014 (2002)
[arXiv:hep-ph/0107144]; \\
A.~H.~Hoang,
arXiv:hep-ph/0412160.

\bibitem{vnrqcd}
M.~E.~Luke, A.~V.~Manohar and I.~Z.~Rothstein,
Phys.\ Rev.\ D {\bf 61}, 074025 (2000)
[arXiv:hep-ph/9910209]; \\
A.~V.~Manohar and I.~W.~Stewart,
Phys.\ Rev.\ D {\bf 62}, 014033 (2000)
[arXiv:hep-ph/9912226]; \\
A.~V.~Manohar and I.~W.~Stewart,
Phys.\ Rev.\ D {\bf 62}, 074015 (2000)
[arXiv:hep-ph/0003032].

\bibitem{rgipNRQCD}
A.~Pineda and A.~Signer,
arXiv:hep-ph/0601185; \\
A.~Pineda and A.~Signer, in preparation.

\bibitem{Beneke:1999qg}
M.~Beneke, A.~Signer and V.~A.~Smirnov,
Phys.\ Lett.\ B {\bf 454}, 137 (1999)
[arXiv:hep-ph/9903260].

\bibitem{c1nnlo}
A.~Czarnecki and K.~Melnikov,
Phys.\ Rev.\ Lett.\  {\bf 80}, 2531 (1998)
[arXiv:hep-ph/9712222]; \\
M.~Beneke, A.~Signer and V.~A.~Smirnov,
Phys.\ Rev.\ Lett.\  {\bf 80}, 2535 (1998)
[arXiv:hep-ph/9712302].

\bibitem{Beneke:1999zr}
M.~Beneke,
arXiv:hep-ph/9911490.

\bibitem{static}
Y.~Schroder,
Phys.\ Lett.\ B {\bf 447}, 321 (1999)
[arXiv:hep-ph/9812205]; \\
M.~Peter,
Phys.\ Rev.\ Lett.\  {\bf 78}, 602 (1997)
[arXiv:hep-ph/9610209]; \\
M.~Peter,
Nucl.\ Phys.\ B {\bf 501}, 471 (1997)
[arXiv:hep-ph/9702245].

\bibitem{shqmc}
C.~W.~Bauer and A.~V.~Manohar,
Phys.\ Rev.\ D {\bf 57}, 337 (1998)
[arXiv:hep-ph/9708306]; \\
E.~Eichten and B.~Hill,
Phys.\ Lett.\ B {\bf 243}, 427 (1990); \\
A.~F.~Falk, B.~Grinstein and M.~E.~Luke,
Nucl.\ Phys.\ B {\bf 357}, 185 (1991); \\
B.~Blok, J.~G.~Korner, D.~Pirjol and J.~C.~Rojas,
Nucl.\ Phys.\ B {\bf 496}, 358 (1997)
[arXiv:hep-ph/9607233].

\bibitem{Pineda:2001ra}
A.~Pineda,
Phys.\ Rev.\ D {\bf 65}, 074007 (2002)
[arXiv:hep-ph/0109117].


\bibitem{Pineda:2001et}
A.~Pineda,
Phys.\ Rev.\ D {\bf 66}, 054022 (2002)
[arXiv:hep-ph/0110216].

\bibitem{c1nnll}
B.~A.~Kniehl, A.~A.~Penin, M.~Steinhauser and V.~A.~Smirnov,
Phys.\ Rev.\ Lett.\  {\bf 90}, 212001 (2003)
[arXiv:hep-ph/0210161]; \\
A.~H.~Hoang,
Phys.\ Rev.\ D {\bf 69}, 034009 (2004)
[arXiv:hep-ph/0307376]; \\
A.~A.~Penin, A.~Pineda, V.~A.~Smirnov and M.~Steinhauser,
Nucl.\ Phys.\ B {\bf 699}, 183 (2004)
[arXiv:hep-ph/0406175].

\bibitem{lognnnlo}
N.~Brambilla, A.~Pineda, J.~Soto and A.~Vairo,
Phys.\ Rev.\ D {\bf 60}, 091502 (1999)
[arXiv:hep-ph/9903355]; \\
B.~A.~Kniehl and A.~A.~Penin,
Nucl.\ Phys.\ B {\bf 563}, 200 (1999)
[arXiv:hep-ph/9907489]; \\
N.~Brambilla, A.~Pineda, J.~Soto and A.~Vairo,
Phys.\ Lett.\ B {\bf 470}, 215 (1999)
[arXiv:hep-ph/9910238]; \\
B.~A.~Kniehl and A.~A.~Penin,
Nucl.\ Phys.\ B {\bf 577}, 197 (2000)
[arXiv:hep-ph/9911414].


\bibitem{Kniehl:2001ju}
B.~A.~Kniehl, A.~A.~Penin, M.~Steinhauser and V.~A.~Smirnov,
Phys.\ Rev.\ D {\bf 65}, 091503 (2002)
[arXiv:hep-ph/0106135].

\bibitem{Beneke:2005hg}
A.~A.~Penin, V.~A.~Smirnov and M.~Steinhauser,
Nucl.\ Phys.\ B {\bf 716}, 303 (2005)
[arXiv:hep-ph/0501042]; \\
M.~Beneke, Y.~Kiyo and K.~Schuller,
Nucl.\ Phys.\ B {\bf 714}, 67 (2005)
[arXiv:hep-ph/0501289].



\bibitem{Hoang:2004tg}
A.~H.~Hoang and C.~J.~Reisser,
Phys.\ Rev.\ D {\bf 71}, 074022 (2005)
[arXiv:hep-ph/0412258].

\bibitem{Chapovsky:2001zt}
A.~P.~Chapovsky, V.~A.~Khoze, A.~Signer and W.~J.~Stirling,
Nucl.\ Phys.\ B {\bf 621}, 257 (2002)
[arXiv:hep-ph/0108190].

\bibitem{upet}
M.~Beneke, A.~P.~Chapovsky, A.~Signer and G.~Zanderighi,
Phys.\ Rev.\ Lett.\  {\bf 93}, 011602 (2004)
[arXiv:hep-ph/0312331]; \\
M.~Beneke, A.~P.~Chapovsky, A.~Signer and G.~Zanderighi,
Nucl.\ Phys.\ B {\bf 686}, 205 (2004)
[arXiv:hep-ph/0401002].

\bibitem{Beneke:2004xd}
M.~Beneke, N.~Kauer, A.~Signer and G.~Zanderighi,
arXiv:hep-ph/0411008.

\bibitem{scet}
C.~W.~Bauer, S.~Fleming and M.~E.~Luke,
Phys.\ Rev.\ D {\bf 63}, 014006 (2001)
[hep-ph/0005275];\\
C.~W. Bauer, S.~Fleming, D.~Pirjol, and I.~W. Stewart,
Phys. Rev. {\bf D63}, 114020 (2001)  [hep-ph/0011336];\\
C.~W. Bauer, D.~Pirjol, and I.~W. Stewart,
Phys. Rev. {\bf D65}, 054022 (2002) [hep-ph/0109045]; \\
M.~Beneke, A.~P.~Chapovsky, M.~Diehl and Th.~Feldmann,
Nucl.\ Phys.\ B {\bf 643}, 431 (2002)  
[hep-ph/0206152].


\end{thebibliography}
\end{document}